\documentclass[epsf]{epl}
\usepackage{epsf}
\begin{document}


\title{Local Spin Glass Order in 1D}

\author{Silvio Franz (1,2) and Giorgio Parisi (3,4)
}
\institute
{ (1) The Abdus Salam
International Centre for Theoretical Physics,\\
Strada Costiera 11, P.O. Box 586, I-34100 Trieste, Italy\\ 
(2) Isaac Newton Institute for Mathematical Sciences\\
20 Clarkson Road, Cambridge, CB3 0EH, U.K.\\
(3) Dipartimento di Fisica, Universit\`a di Roma ``La Sapienza'', \\
P.le A. Moro 2, 00185 Roma, Italy\\
(4) INFM -- CRS SMC, INFN, Universit\`a di Roma ``La Sapienza'', \\
P.le A. Moro 2, 00185 Roma, Italy
}

\date{\today}

\maketitle

\begin{abstract}
We study the behavior of one dimensional Kac spin glasses as function
of the interaction range. We verify by Montecarlo numerical
simulations the crossover from local mean field behavior to global
paramagnetism. We investigate the behavior of correlations and find
that in the low temperature phase correlations grow at a faster rate then 
the interaction range. We completely characterize the growth of correlations 
in the vicinity of the mean-field critical region. 

\end{abstract}

\section{Introduction}
Spin glasses are well understood at the mean field level. After more
then twenty years from the physical clarification of the nature of the
spin glass phase of the Sherrington-Kirkpatrick model \cite{mpv},
recent progress in mathematical physics \cite{guerra,talagrand} is
rapidly leading to a complete mathematical confirmation of the
physical implications of replica symmetry breaking
(RSB). Unfortunately, as soon as one goes beyond mean field, our
ability to make predictions becomes more limited. Large theoretical
efforts devoted to extend the replica symmetry breaking theory to
finite dimensional systems \cite{cirano}, have not led to an unanimous
consensus on the nature of the spin glass phase in finite dimension.
Approximate renormalization schemes \cite{mk} and phenomenological
theories commonly known as droplet models \cite{fh}, suggest that
glassiness in finite dimension could be very different from the mean
field, and indeed much simpler.  Rigorous attempts to describe low
temperature spin glasses are fully compatible both with the replica
symmetry breaking scenario and with droplet like spin glass phases
\cite{ns}. Numerical simulations \cite{numeri} while giving strong
indications that replica symmetry braking might extend down to
dimension three, do not solve the controversy, since it is always
possible to argue that the systems are not large enough, the samples
are not equilibrated etc. It has been recently suggested that the
nature of the low temperature finite D spin glasses can be studied in
an asymptotic expansion around mean field using models with long but
finite interactions of the Kac kind \cite{FT1}. Some progress has
been achieved, proving that for large enough interaction ranges, the
free-energy is close to the mean field. Moreover, if one admits a mild
hypothesis of stochastic stability of the Gibbs metastate w.r.t.  some
random perturbations, mean-field spin glass order holds at least on a
local level: the distribution of overlaps on scales of the order of
the interaction range is close to its mean field limit.  The problem
about the nature of the spin glass phase in finite dimension can be
rephrased as the question if the mean-field order can become
long-range above a finite lower critical dimension. It is clear that
in low enough dimension ($D=1,2$) long range order is not possible and
local mean field order should cross over to paramagnetic behavior on
large scales. Even if one feels that this case is much simpler that
the high dimensional one, a theoretical approach to this cross-over is
at present still to be developed.  In this letter we initiate the
study the crossover from spin glass to paramagnetic behavior in a 1D
spin glass model with variable interactions via Monte Carlo numerical
simulations.

We first verify in explicit simulations the expected property of
genericity of the unperturbed model, and show that local overlaps
approach mean field behavior. Then we investigate the growth of the
correlation length for spin-glass order with the interaction range,
and we find that, analogously to non-disordered models, the
correlation length grows more rapidly than the interaction range.

\section{The model} The model we consider consists in a chain of spins $\sigma_i$
($i=1,...,L$) with periodic boundary conditions, interacting through
the Hamiltonian
\begin{eqnarray}
H[\sigma]=-\sum_{\mu=1}^M J^\mu \sigma_{i^\mu} \sigma_{j^\mu}
\end{eqnarray} 
where $M=(z /2) L$, the indexes $({i^\mu},{j^\mu})$ are chosen 
independently from term to term with uniform probability among
the couples such that $|{i^\mu}-{j^\mu}|\leq R$ for some $R$ while the
$J^\mu$ are i.i.d.r.v. equal to $\pm 1$ with equal probability
\cite{gt}. If $R=L$ the model reduces to the Viana-Bray diluted spin
glass, that for $z>2$ admits a low temperature mean field spin glass
phase \cite{VB}.  The phase diagram of the model is very simple: for
finite $R$, the model is paramagnetic at any positive temperature.  On
the other hand it has been shown \cite{FT3} that in the Kac limit,
$R\to\infty$ after the thermodynamic limit $L\to\infty$, one recovers
the mean-field phase diagram, with a second order phase transition at
a temperature $T_c=1/tanh^{-1}(\sqrt{z})$, below which the system is
in a spin glass phase with full RSB \cite{VB}.

\section{Results} 
In order to characterize the behavior of the system we study the local
overlap between configurations on a scale $R$. We partition the line
$\{1,...,L\}$ into disjoint, contiguous boxes $B_x$, ($x=1,...,L/R$)
of size $R$ and consider the local overlap between spin configurations
$\sigma_i$, and $\tau_i$ as: $q_x(\sigma,\tau)=\frac 1 R\sum_{i \in
B_x} \sigma_i \tau_i$.  A simple generalization of the proof given in
\cite{FT2} shows that if one couples the original Hamiltonian with
suitable finite range random perturbations, one generically has that
the probability distribution of the local overlap, induced by the
Boltzmann distribution on spin configurations $\sigma$ and $\tau$ and
and the quenched random couplings, is close to the overlap
distribution function (ODF) of the infinite range Viana-Bray model for
the same temperature and value of $z$. This has the characteristic
shape of mean-field spin glasses with full RSB, with two delta peaks
at the extremal values $\pm q_{EA}$ and a smooth part in between
\cite{mpv}.  ``Generically'' refers here to the fact that the property
is proven almost everywhere in an interval of values of the couplings
with the perturbations. The stochastic stability property amounts to
say that the case of zero couplings is not a singular exception. In
order to check that indeed this is the case, we simulated the model in
1D for a value of $z=3$, where the mean field critical temperature is
$T_c=1.5186$. In order to equilibrate the system for large samples we
used parallel tempering \cite{pt}. In this way we could reach
interaction ranges $R=256$ for system sizes of $L= 8192$ without
appreciable finite $L$ effects. All quantities we study are averaged
over 100 different samples, and we have checked the stability of the
average, comparing with the average over 50 samples.

\begin{figure}
\begin{center}
\epsfxsize=8cm
\epsffile{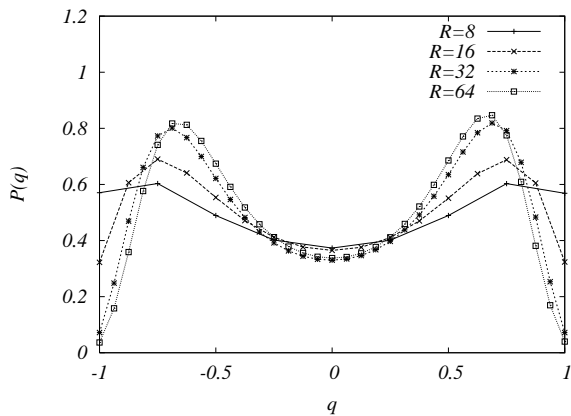} 
\end{center}
  \caption[0]{ Distribution of the overlap on scales $ R$ for
  $T=0.714$ and $R=8,16,32,64$. Increasing $R$ the overlap PDF
  approaches the mean-field distribution.
\label{pR}
}
\end{figure}

\begin{figure}
\begin{center}
\epsfxsize=8cm
\epsffile{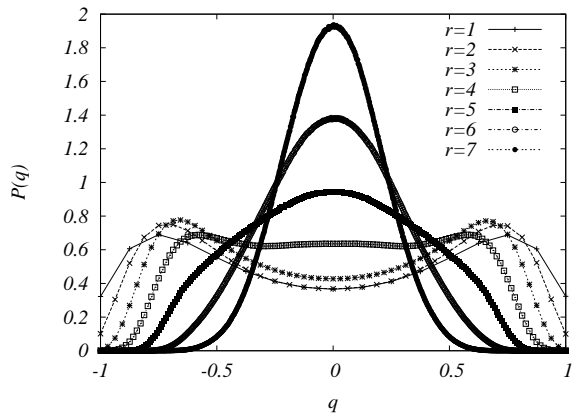} 
\end{center}
  \caption[0]{ Distribution of the overlap on scales $\ell R$ for $T=0.714$ and $R=16$,  $\ell=2^r$ with $r=1,2,3,4,5,6,7$. It is apparent a cross-over from 
mean-field like two peak behavior to paramagnetic Gaussian behavior. 
\label{p16}
}
\end{figure}

In figure \ref{pR} we show the behavior of the function
of $P_R(q)$ for $T= 0.714$ and various values
of $R$. It is apparent that in both cases, increasing the interaction
range, the function $P_R(q)$ approaches the characteristic mean field
shape. This contrast with the behavior of the PDF of the 
global overlap, which for a paramagnet in the thermodynamic limit
has a single delta peak in zero. 
We then study the crossover to paramagnetic behavior of the overlap on
increasing the scale of observation. To this scope, we define overlaps
as before, but on boxes of size $\ell R$ and study the distribution as
a function of $\ell$. In figure \ref{p16} this is done for $R=16$ and
$T= 0.714$, where we see a clear passage from spin-glass behavior at
short lengths to paramagnetic behavior at large scale.

In order to study more
quantitatively this crossover, we considered the
overlap-overlap correlation function
\begin{eqnarray}
C(x) = \langle q_y q_{x+y}\rangle
\end{eqnarray}
where $\langle \cdot\rangle$ denotes average over the Boltzmann and
quenched coupling distribution. Paramagnetic behavior means that this
function should tend to zero at large distance, and in one dimension one 
can expect the behavior:
\begin{eqnarray}
C(x) = C(0) \exp(-x/\xi)
\end{eqnarray}
where $C(0)=\langle q_y^2\rangle$ is the local average of the overlap
square, and $\xi$ is the correlation length of the system, which is
expected to diverge in the low temperature region $T\le T_c$ for 
$R\to\infty$. 

In this letter we concentrate on the critical region where
$\tau=\beta-\beta_c$ is small. In that case one should cross-over to
paramagnetic behavior for finite $R$ to mean field critical behavior
for $R\to\infty$, in which case $\xi\approx \frac{1}{\sqrt{|\tau|}}$
and, for positive $\tau$, $C(0)=\langle q^2\rangle\approx \tau^2$.
Analogously to the well studied case of non-disordered systems
\cite{peli}, the crossover will be described by scaling functions:
\begin{eqnarray}
& &C(0)=\langle q^2\rangle(\tau,R)=\tau^2 g(\tau R^{\alpha})
\\
& &\xi(\tau,R)=\frac{1}{|\tau|^{1/2}} h(\tau R^{\alpha})
\end{eqnarray}
where the properties of the functions $g$ and $h$ for large and small
argument $\tau R^{\alpha}$ should be compatible with the expected
behavior: namely $h$ should go to positive constants for large
negative values of $x$ while it should behave as
$\sqrt{x}$ for small argument to cut-off the mean-field
singularity. Analogously, $g$ should go to a positive constants for
large positive values of $x$, to zero for large negative values and
behave as $\frac{1}{x^2}$ for small $x$.

The value of exponent $\alpha$ can be guessed through the observation
that RSB effects should not affect the critical properties of the
system for $T>T_c$. In that case, one can argue that the fluctuations of the
order parameter are captured by a cubic field theory \cite{cl}
which in the case of the 1D Kac model reads:
\begin{eqnarray} 
F[q]=R \int dz \; \left( \left(\nabla q(z) \right)^2 -\tau q(z)^2 +\lambda
  q(z)^3\right)
\label{cubic}
\end{eqnarray}
where $\lambda$ is a positive constant. Simple scaling analysis predicts
then the value $\alpha=2/5$. This field theory should also in
principle suggest the behaviour of the function $h(x)$ for large negative arguments, 
but we did
not attempt to follow this route.

\begin{figure}
\begin{center}
\epsfxsize=8cm
\epsffile{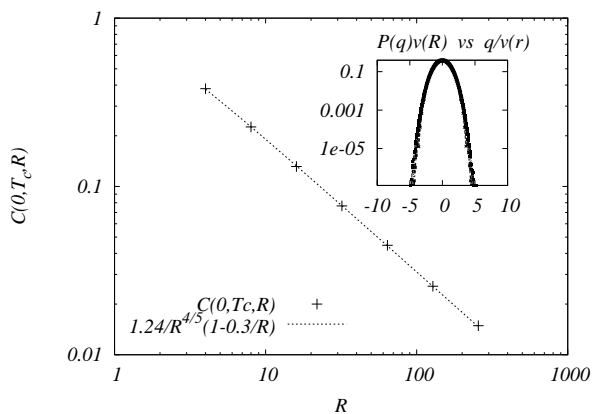} 
\end{center}
  \caption[0]{$C(0)$ as a function of $R$ for $T=T_c$ and best fit of
    the kind $v(R)=\frac{a}{R^{4/5}}\left(1+\frac b R\right)$. In the
    inset, scaling plot of the whole local overlap PDF $P(q)v(R)$
    versus $\frac{q}{v(R)}$.
\label{critico}
}
\end{figure}

\begin{figure}
\begin{center}
\epsfxsize=8cm
\epsffile{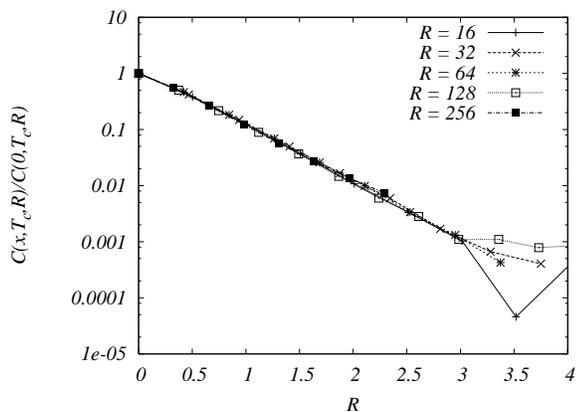} 
\end{center}
  \caption[0]{ Collapse of $C(x)/C(0)=f(x/\xi)$ for $T=T_c$, 
    $R=16,32,64,128,256$. The collapse is obtained for
    $\xi=\frac{R^{1/5}}{1-2.0/R}$.
\label{collapse}
}
\end{figure}

In order to confirm our predictions about the critical exponents and
determine the scaling functions we simulated the model for
$z=3$ and various values of $R$ and temperatures.  We first
investigated the behavior of the overlap correlation function at the
critical temperature $T_c$, in figure \ref{critico} we plot the
behavior of $C(0)$ as a function of $R$ showing that the expected
behavior $C(0)\approx 1/R^{4/5}$ is very well respected: we find small
corrections to scaling that only affect the data points for $R=4$ and
$8$.  In the inset, we plot the scaled function $P(q)$, showing that
the expected scaling holds for the whole probability distribution. 
We next investigated the behavior of the correlation function. In
figure \ref{collapse} we show the collapse of $C(x)/C(0)$ when plotted
as a function of $x/\xi$ with $\xi=\frac{R^{1/5}}{(1-2/R)}$. Again, we
verify our scaling form, assuming small deviations from scaling.
Notice that the assumed exponential form for the correlations is very
well verified for all $x\ge 1$. The same holds at all the temperatures
we looked at.

\begin{figure}
\begin{center}
\epsfxsize=8cm
\epsffile{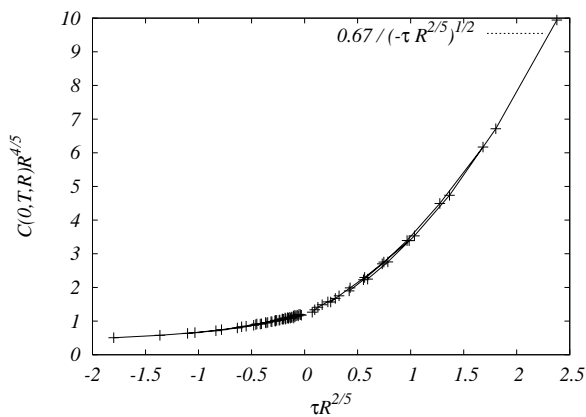} 
\end{center}
  \caption[0]{Scaling of the $C(0,T,R)$ close to $T_c$. In the high
negative argument region we show the fit $\tilde{g}(x)=0.67/\sqrt{-x}$ in
         the region $x<-0.5$, which is indistinguishable from the data.  
\label{scaling-c0}
}
\end{figure}

\begin{figure}
\begin{center}
\epsfxsize=8cm
\epsffile{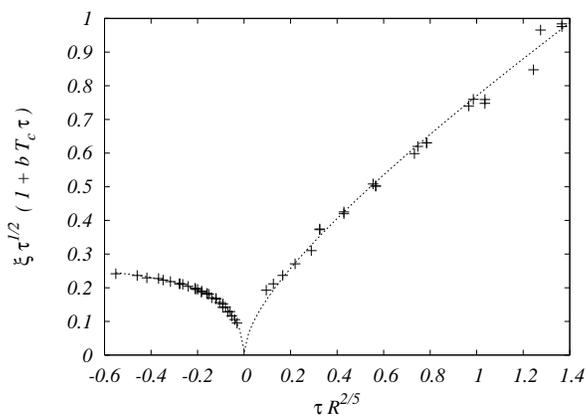} 
\end{center}
  \caption[0]{ Critical scaling of the correlation length.  The curves
are fitted by $h(x)= 0.3 \sqrt{\frac{-x}{0.277-x}}$ for negative $x$ and
by $h(x)=0.42 \sqrt{x}-0.35 x$ for positive x. 
\label{scaling-xi}
}
\end{figure}

We then pass to the task of evaluating the scaling functions.  In figure
\ref{scaling-c0} we plot, for various temperatures, $C(0)R^{4/5}$ as a
function of $\tau R^{2/5}$ which gives the function $\tilde g(x)=x^2 g(x)$.
As expected, $\tilde g(x)$ behaves quadratically for large positive
values of $x$.  The behavior at large negative values can also be
understood, since this is the regime where the effective coupling
constant in the cubic theory should tend to zero and $C(0)\approx
\frac{1}{R\sqrt{|\tau|}}$, leading to $\tilde{g}(x)\approx
\frac{1}{\sqrt{x}}$.  Analogously we can understand the behavior of
the scaling function $h(x)$ (see fig.  \ref{scaling-xi}): for small
$x$, $h(x)$ has the expected square root singularity with a different
prefactor above and below the critical temperature, as it is usual,
while it goes to a constant for large negative values.  The behavior
of the function $h$ for large positive values should describe the
cross-over from critical to low temperature behavior, in particular,
it should determine the behavior of the correlation length in the low
temperature phase. Unfortunately, the data we have, 
though
indicate that a power law behavior $\xi\sim R^\omega$ may persist at
low temperature with an exponent $\omega$ larger then $1/5$, do not
allow a precise determination of the exponent $\omega$ which would
require larger interactions ranges $R$.
\section{Summary} Summarizing we have studied the crossover from paramagnetic to mean
field behavior in a 1D Kac spin-glass model. Our work has a
qualitative aspect, from which we get evidence that stochastic stability
holds at low temperature, and a quantitative aspect where we study the
crossover from paramagnetic to mean-field behavior close to $T_c$. We
characterize this crossover through scaling functions describing the
behavior of the local Edwards-Anderson parameter $C(0)$ and the
correlation length $\xi$.  We find that while in the high temperature
phase the correlation length is, in units of the interaction range
$R$, independent of $R$, in the low temperature phase a dependence on
$R$ sets in.  The correlation length grows as $R^{1/5}$ in units of
$R$ at the critical point, while it grows faster at lower
temperatures. This means that in units of lattice constant the
correlation length grows as $R^{1+1/5}$ or faster.  This result shows
that the rigorous analysis of \cite{FT2} just provides a lower bound to
the size of the regions where local mean-field order holds.  Further
studies will be necessary to make a quantitative analysis deep in the
low temperature region and to go to higher dimension.

{\bf  Acknowledgments}

This work was supported in part by the European Community's Human
Potential programme under contract ``HPRN-CT-2002-00319
STIPCO''. S.F. wishes to thank F.L. Toninelli for useful discussions.

\end{document}